EJSM ORIGINS WORKING GROUP

# EJSM Origins White Document

Recommendations by the Origins Working Group for EJSM Mission – JGO and JEO spacecrafts

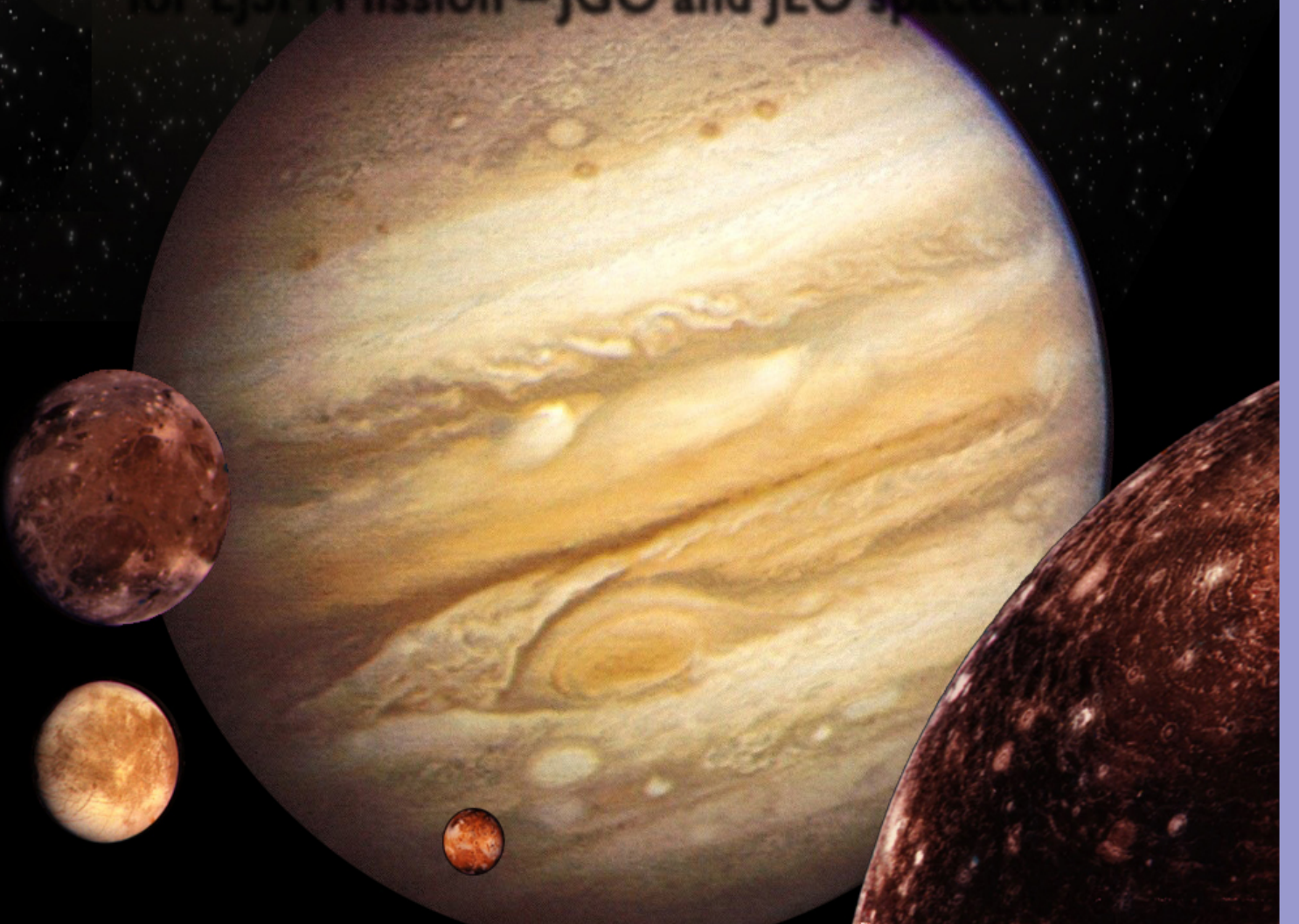

# EJSM Origins White Document

Recommendations by the Origins Working Group for EJSM Mission – JGO and JEO spacecrafts

A. Coradini, D. Gautier, T. Guillot, G. Schubert, B. Moore, D. Turrini and H. J. Waite

# EJSM Origins White Document

A. Coradini, D. Gautier, T. Guillot, B. Moore, G. Schubert, D. Turrini and H. Waite

In the ESA Cosmic Vision 2015-2025 program the investigation of the conditions for planetary formation and that of the evolution of the Solar System were among the leading themes. These two themes, in fact, are strictly interlinked, since a clear vision of the origins of the Solar System cannot be achieved if we cannot discern those features which are due to the secular evolution from those which are primordial legacies of the formation time.

In the context of the EJSM/Laplace proposal for a mission to Jupiter and the Jovian system, the Origins working group tried to identify the measurements to be performed in the Jovian system that could hold clues to unveil the formation histories of Jupiter and the Solar System. As we anticipated, part of the aspects to be investigated is entwined with the study of the evolution of the Solar and Jovian systems.

The identified measures and their scientific domains can be divided into three categories, which will be described in the next sessions.

## *Jupiter and the origins of giant planets*

As concerns the origin of Jupiter, the main interest of the EJSM mission lies in the investigation of Jupiter's composition and internal structure. The internal structure (i.e. the density profile and the mass distribution) of Jupiter holds in fact important clues on the way the planet formed. The study of the Jovian composition and the elemental and isotopic abundances, instead, could supply precious information on the time-scale the formation process acted on and on the environment the planet generated from.

When describing the primordial phases of the origin of the Solar System, the formation of the solar nebula is associated with the collapse of a fragment of an interstellar cloud. This is the only stage which is common to all formation scenarios, since they postulate very diverse histories both in terms of the mechanisms governing the planetary formation than those driving the dissipation of the solar nebula (see e.g. Lunine et al. (2004) and Coradini, Magni & Turrini (2010) and references therein). The main classes of models assume either that a proto-planetary core formed first, and then collected a collapsing envelope of gases and planetesimals (nucleated instability model), or that a gravitational instability in the nebula triggered a local collapse into a proto-planet (ibid). Depending on the process and the time-scale considered, both volatile and refractory materials and heavy elements are delivered to the planet from different sources or, more precisely, from sources at different stages of their evolution. The composition of the circum-solar and circum-Jovian disks and the abundances of the different elements and the chemical species vary, in fact, with time and the disk opacity (see e.g. Dullemond et al., 2006, Bergin et al., 2006, and Coradini, Magni & Turrini, 2010). The relative abundances and the isotopic ratios of the refractory materials and the heavy elements associated to the planetesimals captured by the planet during its accretion also depend on time and the evolution of the disk. The formation region of icy planetesimals in fact is linked to the temperature profile in the disk (see e.g. Jewitt et al., 2006) and therefore depends on time and the disk opacity too. The potential differences in the characteristics of the icy planetesimals would have affected the processes of both planetary formation and surface cratering and they would ultimately influence the volatile content of the Earth and the other terrestrial planets.

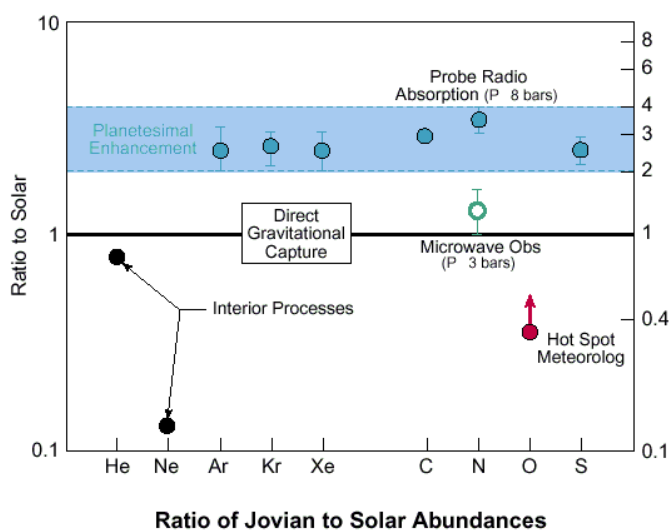

*Figure 1. Elemental abundances (relative to hydrogen) in Jupiter's atmosphere, compared to solar abundances. Updated from Lunine et al. (2004), original version by Owen et al. (1999).*

Jupiter is made mostly of hydrogen and helium (more than 87% by mass), and as such it bears a close resemblance to the Sun. However, the Sun has only 2% of its mass in heavy elements, whereas Jupiter has between 3% and 13% of its mass (Lunine et al., 2004). Moreover, the Galileo probe locally measured abundances of volatile species in excess of solar values (ibid, see Figure 1). The knowledge on the amount and distribution (segregated in a core, or diluted in the envelope) of both heavy elements and volatile species is crucial to understand Jupiter's formation and to constrain the thermodynamic conditions in the early Solar Nebula. The knowledge on the existence and the structure of the core, in turn, would help solve the controversy between the two major scenarios for the formation of giant planets. At present, the information at our disposal on the giant planets in the Solar System argues in favour of the nucleated instability scenario (Lunine et al., 2004). In such a scenario the planets should first form a solid core of the order of 10 Earth masses that would act as the accretion centre for the gas of the Solar Nebula, which will constitute the dominant fraction of their final masses. The limited lifetime of the Solar Nebula, which has been constrained to about 10 Myr through astronomical observations of circumstellar disks around near-by stars (Meyer et al., 2006), poses a strict constrain to the formation time of the planetary core and the accumulation of the gas. Yet, this upper limit to the formation time is also the most detailed information we currently possess on the temporal evolution of such processes. Knowing the mass and the size of the core would then allow for a clearer view of the events which took place at such remote times, therefore investigating the internal structure of Jupiter is a scientific priority of level 1 for the *Origins* theme.

To date, the internal structure of the giant planets has been investigated through the study of their gravitational momenta $J_n$. This method, while well tested, is much more sensitive to the distribution of matter in the outer half of the planetary radius than to that of the inner half and therefore of the planetary core (see Figure 2). Moreover, the NASA mission JUNO will perform a detailed investigation of the gravitational momenta from a more favourable (i.e. polar) orbit before the arrival of EJSM (see e.g. Blanc et al., 2009). Estimates of the possible contribution of EJSM mission to the evaluation of the Jovian gravitational momenta indicate that the mission will not be able to improve JUNO's measurements. In the context of the EJSM/Laplace mission proposal, therefore, we strongly recommend the study of possible alternative solutions to investigate the interior of Jupiter. To this regard, the Origins working group agrees with the response of ESA's evaluation of the declarations of interest for instruments on-board EJSM and support the further study of the Doppler Spectro-Imager instrument, included in the original Laplace proposal (see e.g. Blanc et al., 2009, and references therein), as a potential new way to approach the investigation of the interior of giant planets.

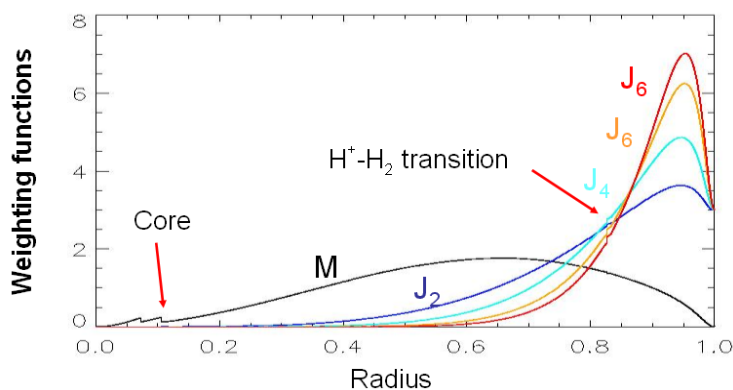

*Figure 2. Weighting functions as a function of Jupiter's radius for different gravitational momenta $J_{2n}$ (colour lines). Black line: Jupiter's mass as a function of the radius. (Guillot, personal communication).*

Another important class of measurements to be done on Jupiter are the elemental and chemical abundances of the key species present in its atmosphere. Owen and Encrenaz (2003, 2006) have suggested that the observed enrichment of heavy elements in Jupiter can be explained only by considering the abundances and isotope ratios. Their original assumption is that the Jovian atmosphere can be considered as a mixture of volatiles lost from the original core, from the solar-composition gases captured further and from icy planetesimals captured and dissolved in the atmosphere. As a complement to the study of the internal structure, therefore, it is important to analyze, through remote sensing, the spectral features of the gases present in the atmosphere of Jupiter since they can be used to determine their relative abundances. The different scenarios for the formation of Jupiter, in fact, predict very diverse relative abundances for the different chemical species (see e.g. Lunine et al., 2004, and references therein). If Jupiter's core is of about 10 Earth masses and the rest of the gas trapped by this core has a solar composition then, the mass of Jupiter being about 318 Earth masses, the overall mass in refractory elements should be about 16 Earth masses. This implies that the global abundance of heavy elements in Jupiter is enriched by a factor of 3 with respect to the solar one (ibid). Again, the Jovian superabundance with respect to the Sun seems to support the nucleated instability scenario (ibid). Therefore, the study of the abundances of different elements and molecules generated in the Jovian reducing atmosphere is a key method to shed some light on the origin of Jupiter. In particular, the He/H ratio gives information on how much helium has been dissolved in metallic hydrogen in the interiors of the planets, constraining the models of internal structure and evolution. The Ne/H ratio, instead, supply information on the possible temperature of planetesimals contributing to the core formation: to trap Ne, in fact, they must have been colder than 15°K (see e.g. De Pater & Lissauer, 2001). Finally, the noble gases Ar, Kr, Xe could all exhibit the same enrichment factor found for $CH_4$ if these gases were trapped by simple freezing or adsorption in icy planetesimals at temperatures below 25°K (ibid). On the other hand, if they were trapped in clathrate hydrates, a variety of different ratios could ensue. As a consequence, the measurements to be supported are those in NIR (where most of the relevant molecules have diagnostic signatures) and MID IR, particularly in the range between 4 and 6 microns where thermal emission dominates over reflected sunlight and the Jovian atmosphere is characterized by a "window" of low opacity which allows one to probe up to a depth of 5 to 8 bars (Taylor et al., 2004). Such set of measurements represents a scientific priority of level 2 for the *Origins* theme. The cost/benefit ratio of extending the spectral range up to 8 microns for the study of the Jovian atmosphere is presently under evaluation (see e.g. Grassi et al., 2009) yet preliminary estimations locate the best trade-off between instrument performances and information gain at about 6 microns (ibid).

## *Inner Jovian system: regular satellites*

While JUNO and EJSM missions aim at the same goal through different approaches as concerns the investigation of the Jovian internal structure, the design of the EJSM mission has been optimized for the study of the regular satellites (Blanc et al., 2009).

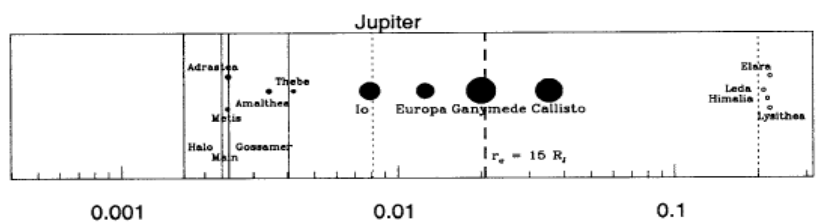

*Figure 3. Jovian satellite systems with the distance scale in terms of the planet's Hill radius (about 750 $R_J$). Planetary rings are denoted by solid lines and are labelled when possible. The centrifugal radius is denoted by a bold dashed line. Figure adapted from Mosqueira & Estrada (2003).*

The main goal is to understand the relationship between the formation of the Galilean moons (see Figure 3) and the origin of Jupiter. The physics and chemistry of the Galilean (regular) satellites can, in fact, be directly related to a physical process similar to the one that, at a larger scale, led to the formation of the planets (see e.g. Coradini, Magni & Turrini, 2010, and references therein). Such processes are regular, and they are physically and chemically continuous, thus the observation of some key parameters in the current Jovian system can help in constraining backwards the conditions in which the regular satellites system formed; particularly the initial conditions of the Solar Nebula and the Jovian sub-nebula, a disk of gas and dust surrounding Jupiter after its formation. This goal can be accomplished by measuring the stable isotopes of C, H, O, and N in the major volatile species ($H_2O$, $CH_4$, $NH_3$, CO, $N_2$, $CO_2$, $SO_2$, etc.). The measurement of the D/H ratio in $H_2O$ and $CH_4$ is particularly important to determine the temperature at the time of the condensation processes of the satellites (see e.g. De Pater & Lissauer, 2001). In addition, to understand the origin and delivery of the volatiles it is important to measure/constrain the content in noble gases (Ne, Ar, Kr, Xe) of the Galilean satellites (ibid). Finally, the evolution of geochemical processes in the satellites can be constrained by measuring the detailed organic distribution of volatile material, e.g. alkenes versus nitriles versus aldehydes versus alcohols. Such information could be achieved through the use of visual and near-infrared spectroscopy to study the surface composition combined with mass spectrometry of the exospheres of the satellites. Spectroscopic information is needed with higher spectral resolution than the one supplied by Galileo, both in the IR and VIS ranges (i.e. a wider global spectral range with respect to past missions), in order to discriminate non-water compounds and their degree of mixing with water ice. The inclusion of mapping capabilities into the necessary VIS-NIR spectrometer would greatly help any data fusion with camera images, aimed to localize spectral units possibly related to the geophysics of the satellites. The minimum spectral range that is requested is, in the IR, 1-5 micrometers. Typically, instruments working in this range have a spectral resolution of the order of 9-10 nm: for what it concerns the *Origins* theme, it will be important to evaluate the feasibility of designing spectrometers with higher (i.e. of the order of 5 nm) spectral resolution around 3 micrometers to better discriminate the presence, pure or as clathrates, of organic compounds. Such measurements represent a scientific priority of level 1 for the *Origins* theme.

Thanks to its two-spacecraft design, EJSM will investigate in details both Europa and Ganymede and will supply data on Io and Callisto during its cruise (Blanc et al., 2009). Of the four Galilean moons, Ganymede and Callisto show the major abundance of water ice (revealed by their bulk density), reflecting the temperature distribution in the Jovian sub-nebula (see e.g. Lunine et al., 2004, and Coradini, Magni & Turrini, 2010). This statement is based on the Galileo's observations of the surface composition and on the gravity measurements from which the internal composition is inferred (Schubert et al., 2004, and references therein). Gravity measurements will be strongly improved for Ganymede and Europa, given the possibility to perform radio-science observations from quasi-polar orbits. Moreover, the limited dataset supplied by Galileo on Callisto (ibid) will greatly benefit from the possibility to perform non-equatorial fly-bys of the satellite. These measurements will permit to better constraint the internal structures of the Galilean satellites, which are the result of their complex thermal histories that are in turn related to the strength of the energy sources that were present during their evolutions. In particular, if the need of the presence of short-lived radioactive elements is confirmed, this will pose important constraints on the formation time and timescale of the satellites (see Castillo-Rogez et al., 2007, for analogous considerations for the Saturnian satellites and Barr and Canup, 2008, for a discussion of the possible implications for Callisto). As a consequence, such measurements are a scientific priority of level 1 for the *Origins* theme. An independent indication of the presence and composition of inner conductive layers comes from the measure of the intrinsic magnetic field, as well as of the one induced by the interaction with the Jovian radiation belts (see e.g. Kivelson et al., 2004). The study of the characteristics of the intrinsic and induced magnetic fields will permit to better constraint the internal composition of the satellites and is a scientific priority of level 2 for the *Origins* theme.

The density measurement shall also be combined with the information of the surface composition, deduced from the analysis of spectrometers present on EJSM. Because their orbits set outwards of the orbits of Io and Europa, Ganymede and Callisto are less affected by the interactions with the Jovian magnetosphere. In this sense, the surfaces of Ganymede and, particularly, Callisto, are believed to be less altered and thus more representative of the original composition of the sub-nebula; therefore studying these bodies by a multiple close flybys campaign as well as from a dedicated orbit, and through an aimed instrumentation, is of great interest in addressing the science theme related to the origins of the Jupiter system. The measurements to be performed to probe the internal structures of the icy Galilean satellites Ganymede and Callisto are thus basically gravity and magnetic fields measurements and radio science through X and Ka bands, to be combined with the previously suggested spectroscopic measurements.

The investigation of the interior structure can be complemented by the study of the cratering history of the satellites, which is important also from the point of view of understanding the history of the Solar System. Crater counting can provide important information not only to determine the age of the surfaces of the satellites but also to test the reality and the characteristics of the Late Heavy Bombardment, which has been suggested to be triggered in the Solar System by the combined effects of the migration of the giant planets and their interactions with the residual planetesimal disk (Tsiganis et al., 2005). Finding evidences of such an increased flux of bodies crossing the Jovian region can provide us fundamental clues on the chronology of the formation and evolution of the Solar System and is a scientific priority of level 1 for the *Origins* theme. Therefore, we support the integration of the previously indicated measurements with high resolution imaging of the surfaces of the satellites to allow for a improved crater counting respect to the one permitted by Galileo mission.

## *Outer Jovian system: irregular satellites*

The exploration of the outer Jovian system was one of the goals in the original Laplace proposal (Blanc et al., 2009). Extending up to the boundaries of the stable region of the Jovian Hill's lobe and inhabited by the irregular satellites (see Figure 3), a population of asteroid-like or Centaur-like objects, the outer Jovian system is an interesting region for what it concerns the study of the origin and the evolution of the Solar and Jovian systems.

Irregular satellites, in fact, have long been recognised as celestial bodies which are alien to the satellite systems they belong to (see e.g. Sheppard, 2006, and Jewitt & Haghighipour, 2007, and references therein). Depending on the scenario advocated to explain their origins, irregular satellites are captured either during the last phases of the formation of the giant planets or after the giant planets fully formed, while the outer Solar System was shaping itself to its present configuration (Jewitt & Haghighipour 2007, and references therein; Nesvorny, Vokrouhlicky & Morbidelli, 2007; Turrini, Marzari & Tosi, 2009). While the observational data hint to a similar efficiency in capturing the irregular satellites for all four giant planets (Jewitt & Sheppard, 2005), theoretical and dynamical considerations seem to suggest that the Jovian ones originated differently from those of Saturn, Uranus and Neptune (Nesvorny, Vokrouhlicky & Morbidelli, 2007). Therefore, the question as to whether one or more capture mechanisms acted in the outer Solar System is still open to debate. Yet, it is clear that the origins of the irregular satellites are linked to some of the main evolutionary phases of the outer Solar System and their study can help us to probe its history.

Moreover, irregular satellites are interesting also due to their possible link with regular satellites. Dust production and transfer processes from the irregular satellites to the regular ones were suggested to take place in the Saturnian system, and by analogy in all the satellite systems of the giant planets, in order to explain the dichotomous appearance of Iapetus (see e.g. Tosi et al., 2010, and references therein). In recent years, a number of observational data argued in favour of this scenario, i.e. the enhanced dust population in the outer Jovian system measured by Galileo mission (Krivov et al., 2002), the spectral data on Phoebe and the dark material on Iapetus supplied by Cassini mission (see e.g. Tosi et al., 2010, and references therein) and the discovery of a new Saturnian dust ring spanning over the orbital region of the irregular satellites (Verbiscer et al. 2009). The study of such mass transfer phenomenon would shed new light on the evolution of the Jovian system and therefore help in interpreting its primordial features.

Observing the irregular satellites is a difficult task due to their small size and the vast region of space they populate and the preliminary determination of the orbital tours of the JEO and JGO spacecrafts would not allow for fly-bys of any Jovian irregular satellite. However, we strongly argue in favour of keeping open the possibility of a serendipitous encounter during the crossing of the outer Jovian system in future refinements of the scientific goals of the mission and of the orbital tour of the spacecrafts. The measurements to be performed to investigate the origin and the nature of the irregular satellites in the ideal case would be high resolution imaging and visual-near infrared spectroscopy (with mapping capabilities) of the satellite surfaces during close fly-bys like the one performed by Cassini on Phoebe. If close fly-bys of the irregular satellites should prove unfeasible, we advise to evaluate the feasibility of performing full-disks observations of as many irregular satellites as possible. Another potentially interesting measure would be the D/H ratio of the irregular satellites, which could represent a probe for the formation region of their parent bodies (see e.g. Horner et al., 2008). Such measurement could be carried on with sub-millimetre observations of the irregular satellites and/or through mass spectrometry and infrared spectroscopy, both of which would require close fly-bys of the targets. Given the difficulties, mainly due to orbital constrains, such measurements represent a scientific priority of level 2 for the *Origins* theme.

Independently from the direct investigation of the irregular satellites, we advise to refine the measurements of Galileo mission by sampling the dust population during the cruise towards the inner Jovian system and studying both its dynamical and compositional features. Again, such measurements represent a scientific priority of level 2 for the *Origins* theme. Spectral and radar data on the dark material on Callisto, like the ones which should be supplied by the measurements targeting the regular satellites, could also help completing the dataset to probe its origin.

## *Synergy between JEO and JGO*

The potential synergies between JGO and JEO are related to the investigation of the surface features and composition of the Galilean satellites to probe the thermal and chemical features of the circum-Jovian and the dynamical history of the outer Solar System. In particular, JEO will permit to complement the dataset supplied by JGO for Ganymede and Callisto with information on Europa and Io. The scientific goals interested by the synergy between JGO and JEO are those labelled as A, B, D, G and H in the table in the **Requirements** section.

## *Requirements*

Given the previous considerations, we can define a series of key measurements to be performed by the EJSM JGO spacecraft, both separately and in conjunction with the JEO spacecraft. In particular, it will be extremely important to combine the information on the satellite composition gathered by JEO and JGO. In what follows we have summarized the critical measurements ranked in terms of priorities.

| Goal/Priority | Measure | Subject | Instrument or Technique | Reference Payload | Scientific Goals | Requirements | Bibliography |
|---|---|---|---|---|---|---|---|
| **A/1: Satellite Origins** | **Gravity fields of the satellites** | Callisto, Ganymede, other satellites | Radio Science Transponder-Receiver | JRST | Measurement of gravity field allows determining the internal structure, which in turn allows estimating the degree of differentiation and constrains its formation time and timescale. | Non-equatorial fly-bys of Callisto (all Galileo's ones were nearly equatorial) | Barr & Canup (2008,2010), Schubert et al. (2004) |
| **A/1: Satellite Origins** | **Shape measurements** | Callisto, Ganymede, other satellites | Imaging | MRC,WAC | Search for fossil bulge/non-hydrostatic shape on Callisto and verification of hydrostatic equilibrium assumption for Ganymede | | Blanc et al. (2009) |
| **B/1: Planetary Nebula Origin** | **Composition of the satellites** | Noble Gases (Ar, Kr, Xe, Ne) | Mass Spectrometry | PP (Particle Package) | Estimate of the temperature profile of the Jovian subnebula | | De Pater & Lissauer (2001); Coradini, Magni, Turrini (2010) |
| **B/1: Planetary Nebula Origin** | **Composition of the satellites** | Ices ($H_2O,CH_4,CO,CO_2,N_2$) | IR Spectroscopy | VIRHIS | Estimate of the volatiles and the temperature profile in the Jovian subnebula | | De Pater & Lissauer (2001); Coradini, Magni, Turrini (2010) |
| **B/1: Planetary Nebula Origin** | **Composition of the satellites** | Clathrates ($CH_4,CO,N_2,CH_4,NH_3$) | IR Spectroscopy | VIRHIS | Estimate of the volatiles and the temperature profile in the Jovian subnebula | Resolution should be high enough (i.e. about 5 nm) to discern clathrates from ices | De Pater & Lissauer (2001); Coradini, Magni, Turrini (2010) |
| **B/1: Planetary Nebula Origin** | **Composition of the satellites** | Minerals & Refractory material | IR Spectroscopy | VIRHIS | Estimate of the planetesimal/satellitesimal composition in the Jovian subnebula | Global coverage of the Galilean satellites | De Pater & Lissauer (2001); Coradini, Magni, Turrini (2010) |

| C/1: Satellite Thermal Evolution | Composition of the satellites | Decay product of long-lived radionuclides (e.g. 40-Ar) | Mass Spectrometry | PP (Particle Package) | Investigate energy sources and degree of out-gassing of the satellites | | De Pater & Lissauer (2001); Coradini, Magni, Turrini (2010) |
|---|---|---|---|---|---|---|---|
| C/2: Satellite Thermal Evolution | Internal structure of the satellites | Presence of sub-surface liquid, conductive layers | Magnetometer | MAG | Identify the existence and extension of sub-surface liquid, conductive layers | | Kivelson et al. (2004) |
| D/1: Pristinity of Planetary Nebula's Material | Isotopic abundances of the satellites | D/H | Mass Spectrometry | PP (Particle Package) | Measurement of D/H and other isotopes enrichment supplies information on the temperature profile in the Jovian subnebula and the origin of the nebular material | Constrains depends on the molecule to be investigated: DH/H2 poses no resolution constrain, while discerning C(13)H4 from C(12)H3D requires high resolution (up to about 4000) | De Pater & Lissauer (2001); Horner et al. (2008); Coradini, Magni, Turrini (2010) |
| E/1: Jupiter's Formation | Internal structure of Jupiter | Jupiter's oscillations | Helioseismology | DSI | Knowledge of Jupiter's density profile allows the determination of the existence of the central core, which supplies information on Jupiter's formation process | No previous measurement of this kind. LAPLACE's goals aimed to a noise level of 0.3 cm/s for the detection of oscillations up to mode l=50, and a precision on the frequency measurements better than 0.5 µHz. | Schmider, Fossat & Mosser (1991), Blanc et al. (2009) |

EJSM Origins White Document| | | | | | | | |
|---|---|---|---|---|---|---|---|
| **E/2: Boundary Conditions on Jupiter's Evolution** | **Jupiter's atmospheric composition** | Volatile species (H20, NH3) and High-Z elements | IR Spectroscopy | VIRHIS | Measurement of the enrichment in High-Z material respect to solar composition provides information of the formation process & environment of Jupiter | Evaluate extension of IR range up to 6 micron to improve accuracy | Grassi (2009) |
| **F/2: Link with Protosolar Nebula** | **Composition of the irregular satellites** | Ices (H2O,CH4,CO,CO2,N2) | IR Spectroscopy | VIRHIS | Gather information on their formation region(s) and/or their parent population(s) | Full disk observation or fly-by | Sheppard (2006) |
| **F/2: Link with Protosolar Nebula** | **Composition of the irregular satellites** | Minerals & Refractory material | IR Spectroscopy | VIRHIS | Gather information on their formation region(s) and/or their parent population(s) | Full disk observation or fly-by | Sheppard (2006) |
| **F/2: Link with Protosolar Nebula** | **Isotopic abundances of the irregular satellites** | D/H, C, O, N isotopic ratios | Mass Spectrometry | PP (Particle Package) | Gather information on their formation region(s) and/or their parent population(s) | Close fly-by of irregular satellites are required | Horner et al. (2008) |
| **G/2: Satellite Contamination** | **Dust sampling in the Jovian system** | Dust orbital and compositional features | Dust Analyzer | None | Investigate mass transfer processes between Jovian satellites (also irregular ones) | Inclusion of compositional analysis, improve orbital determination | Krivov et al. (2002), Tosi et al. (2010), Messenger et al. (2003) |
| **H/1: Dating the Satellite Surfaces** | **Imaging of the Galilean and small satellites** | Crater counting | Imaging | HRC,MRC | Dating of the surfaces through crater counting and investigation of the history of the outer Solar System | Improve resolution of global coverage and extend crater counting to diameters < 10 km (about 1 km) | Schenk et al. (2004) |